\documentclass[conference,a4paper]{IEEEtran}
\usepackage{xcolor}
\usepackage{balance}
\usepackage{array}

\ifCLASSINFOpdf
\usepackage[pdftex]{graphicx}
\else
\usepackage[dvips]{graphicx}
\fi
\ifCLASSOPTIONcompsoc
\usepackage[caption=false,font=normalsize,labelfont=sf,textfont=sf]{subfig}
\else
\usepackage[caption=false,font=footnotesize]{subfig}
\fi
\begin{document}
		%
		\title{Mid-band Propagation Measurements in Industrial Environments}

		\author{\IEEEauthorblockN{
				Juha-Matti Runtti\IEEEauthorrefmark{1},   
				Usman Virk\IEEEauthorrefmark{1},   
				Pekka Ky\"{o}sti\IEEEauthorrefmark{2}      
				Lassi Hentil\"{a}\IEEEauthorrefmark{1},    
				Jukka Kyr\"{o}l\"{a}inen\IEEEauthorrefmark{1}      
				and Fengchun Zhang\IEEEauthorrefmark{3}      
			}                                     
			\IEEEauthorblockA{\IEEEauthorrefmark{1}
				Keysight Technologies, Oulu, Finland, juha-matti.runtti@keysight.com}
			\IEEEauthorblockA{\IEEEauthorrefmark{2}
				Keysight Technologies \& University of Oulu, Oulu, Finland, pekka.kyosti@keysight.com}
			\IEEEauthorblockA{\IEEEauthorrefmark{3}
				Aalborg University, Aalborg, Denmark, fz@es.aau.dk}
		}
		


		\maketitle
		
		\begin{abstract}
			6G radio access architecture is envisioned to contain a network of short-range in-X subnetworks with enhanced capabilities to provide efficient and reliable wireless connectivity. Short-range communications in industrial environments are actively researched at the so-called mid-bands or FR3, e.g., in the EU SNS JU 6G-SHINE project. In this paper, we analyze omni-directional radio channel measurements at 10--12 GHz frequency band to estimate large-scale channel characteristics including power-delay profile, delay spread, K-factor, and pathloss for 254 radio links measured in the Industrial Production Lab at Aalborg University, Denmark. Moreover, we perform a comparison of estimated parameters with those of the 3GPP Indoor Factory channel model.
		\end{abstract}
		
		\vskip0.5\baselineskip
		
		%
		
		\section{Introduction}
		6G wireless technologies are poised to revolutionize communication, offering unprecedented connectivity through advanced features such as ultra-fast data rates, low latencies, utilization of new frequency spectrum, and AI-driven networks. The 6G radio access network architecture is envisaged as an umbrella network that coherently combines and interconnects several in-X subnetworks, such as macro, micro, pico, non-terrestrial, drones and unmanned aerial vehicle (UAV), and private networks \cite{Uusitalo2021}. 
		
		The in-X subnetworks are short-range low-power radio cells at the edge of the 6G network to support localized and seamless wireless connectivity. These can be configured in small-scale entities such as robots, production modules, vehicles, and immersive classrooms to avoid wired infrastructure for communication services that have specific data rates, complexity, latency, and reliability requirements \cite{6GSHINE}. It is noteworthy that the subnetworks conceptualize dynamic and trusted local connectivity zones defined in the ITU vision for the future development of International Mobile Telecommunications (IMT) for 2020 and beyond \cite{IMT2020}, and they are also included in the European vision for the 6G Network Ecosystem \cite{EU6GVision}.
		
		Another increasing trend in the telecommunication field is towards the use of frequency bands between current FR1 and FR2 frequency ranges, also termed mid-band or FR3, i.e., 7-24\,GHz. Especially, the lower edge of the mid-band is appealing to the industry. The third intersecting area includes radio applications, such as sensing, that are of particular interest to factories and other production environments. These trends have motivated propagation measurements and channel modeling activities within the European 6G-SHINE (6G Short Range Extreme Communication in Entities) project \cite{6GSHINE}.
		
		In this work, we investigate the radio channel characteristics in a production lab that resembles a typical industrial manufacturing hall. From the electromagnetic wave propagation point of view, special geometrical features of such environments include various irregular metallic surfaces and structures, high ceiling height, and numerous moving objects, such as robots. The research is done by conducting radio propagation measurements with a frequency domain channel sounder, estimating large-scale parameters of the multipath radio propagation channel, and evaluating statistics for these parameters. The main target is to gather an understanding of the channel using empirical methods. We also aim to compare the channel statistics with the 3GPP standard channel model \cite{3GPP38901} for the Indoor Factory (InF) scenario. The performed measurements do not capture temporal variation since only static RX locations were measured. Furthermore, the angular information is not available from this measurement data set because omni-directional antennas were used.
		
		\section{Radio Channel Measurements}
		
		\subsection{Measurement Scenario}
\begin{figure}
			\includegraphics[width=\linewidth]{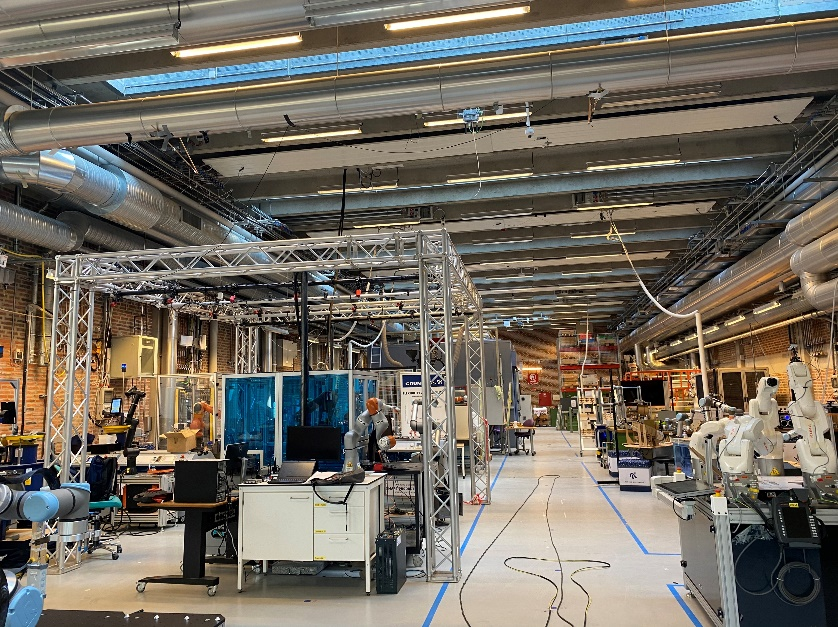}
			\centering
			\caption{A photograph from the measured environment in the Industrial Production Lab at Aalborg University.}
			\label{fig:EnvPhoto}
		\end{figure}

		The channel measurements were undertaken in the Industrial Production Lab at Aalborg University, Denmark. An overview photo of the environment is given in Fig. \ref{fig:EnvPhoto}. The measurement location was a typical indoor factory with large amounts of clutter with different kinds of instruments and metallic production equipment providing significant propagation effects for the channel measurements. There are no solid partition walls within the lab but empty corridors are left between various metallic structures, as can be observed from the photo. Metallic ventilation pipes located horizontally high on walls as well as vertically oriented cylindrical frame structures are expected to provide strong reflections in many directions both in elevation and azimuth. 
		
\begin{figure}
			\includegraphics[scale = 0.4]{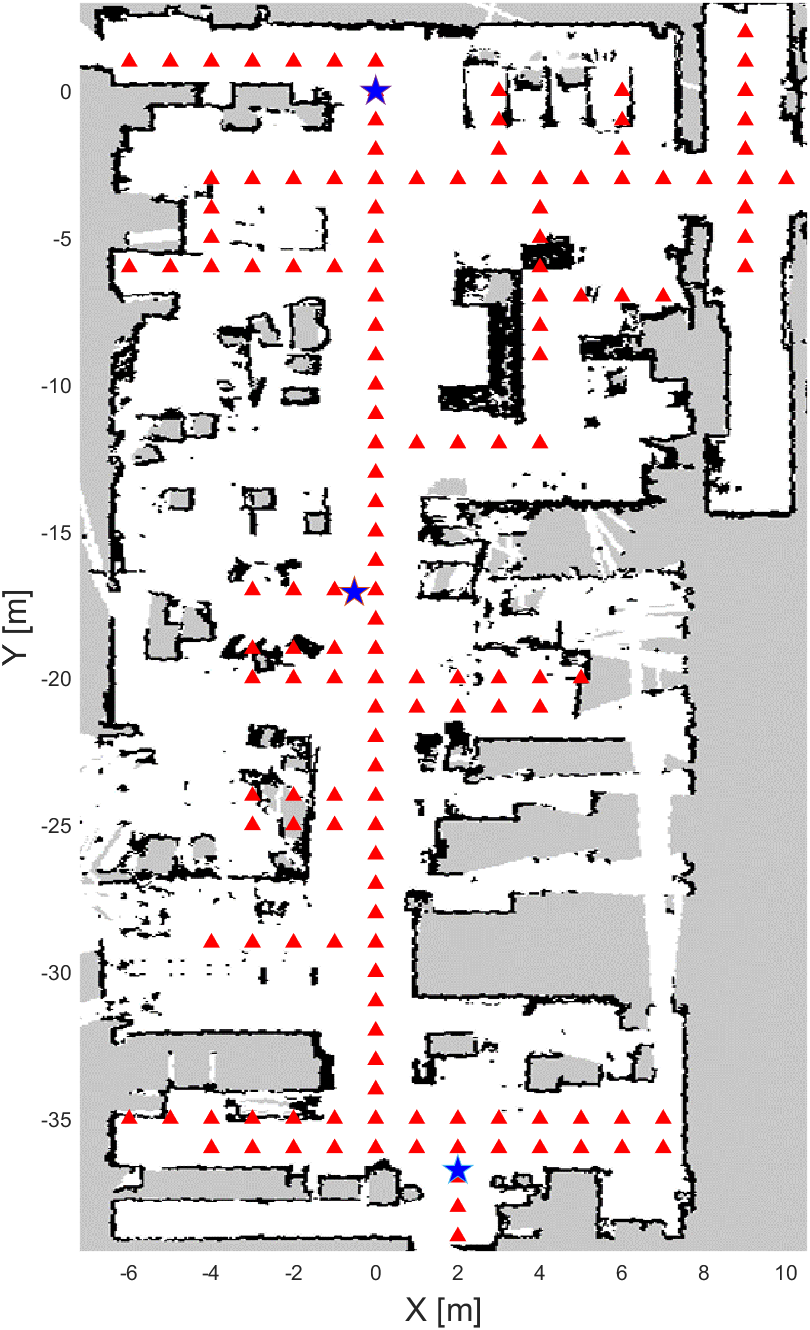}
			\centering
			\caption{Measurements map depicting RX and TX positions. Blue stars and red triangles denote transmitters and receivers, respectively.}
			\label{fig:all_RX_TX_positions}
		\end{figure}

		Measurements included three positions for the TX antenna and 115 positions for the RX antenna. The measurement layout is shown in Fig. \ref{fig:all_RX_TX_positions}, where blue pentagrams depict the transmitter locations and red triangles depict the receiver locations. The layout indicates dimensions of approximately 41~m $\times$ 17~m for the length and width of the hall.
		
		\subsection{Measurement System and Setup}
\begin{figure}
			\textsf{\includegraphics[bb=220bp 220bp 610bp 435bp,clip,scale=0.56]{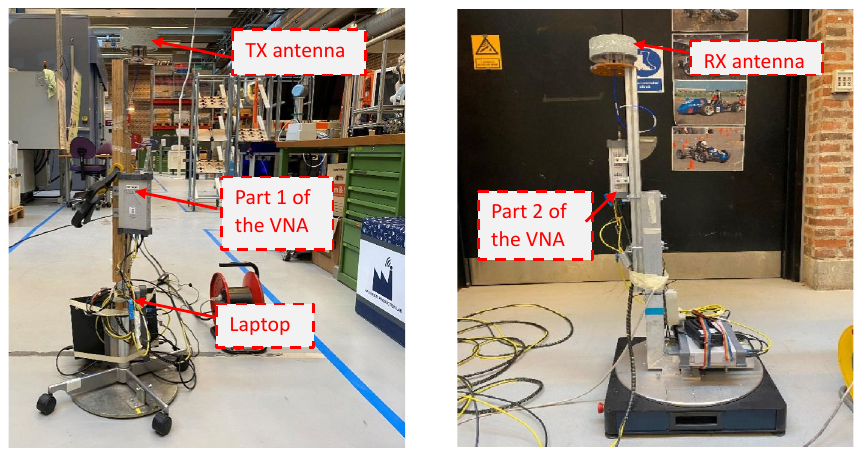}}
			\centering
			\caption{A photograph of measurement system equipment, TX on the left and RX on the right.}
			\label{fig:measurement_system}
		\end{figure}

		A photograph of the measurement system is shown in Fig. \ref{fig:measurement_system}. It is composed of two identical bi-conical antennas (INFO SZ-2003000) used as the transmitter (TX) and the receiver (RX) antenna, Anritsu ShockLine MS46131A-043 \cite{Anritsu} vector network analyzer (VNA) employed to collect the channel frequency response between the TX and the RX antenna ports, and a laptop used to launch the measurement scripts and store the data. VNA parts in both TX and RX are synchronized by Anritsu PhaseLync\texttrademark{} technology.
		
		The antennas covered the frequency range of 2 -- 30~GHz with vertical polarization and an antenna gain of 3.2 dB at 11 GHz. The VNA was configured to sweep frequencies from 10 to 12 GHz with 2001 points. For TX0, the RX antenna was moved to 115 positions. For TX1, there were 89 RX positions, and for the TX2 there were 50 RX positions.
		
		\section{Data-Processing and Analysis}
		
		\subsection{Denoising Measurement Data}
		Identification of true propagation paths and removal of noise samples and possible artifacts is a critical operation in the signal analysis of most empirical propagation studies. In the following, we briefly explain how this was done by setting a noise-cutting power threshold, removal of delay components in the long tail of impulse responses, and identification of the first time of arrival and removing delay bins before that. An extra challenge is the detection of LOS/NLOS conditions since the lack of interior walls and the existence of very complex structures in the hall prevented direct layout-based categorization.
		
\begin{figure}
			\includegraphics[width=\linewidth]{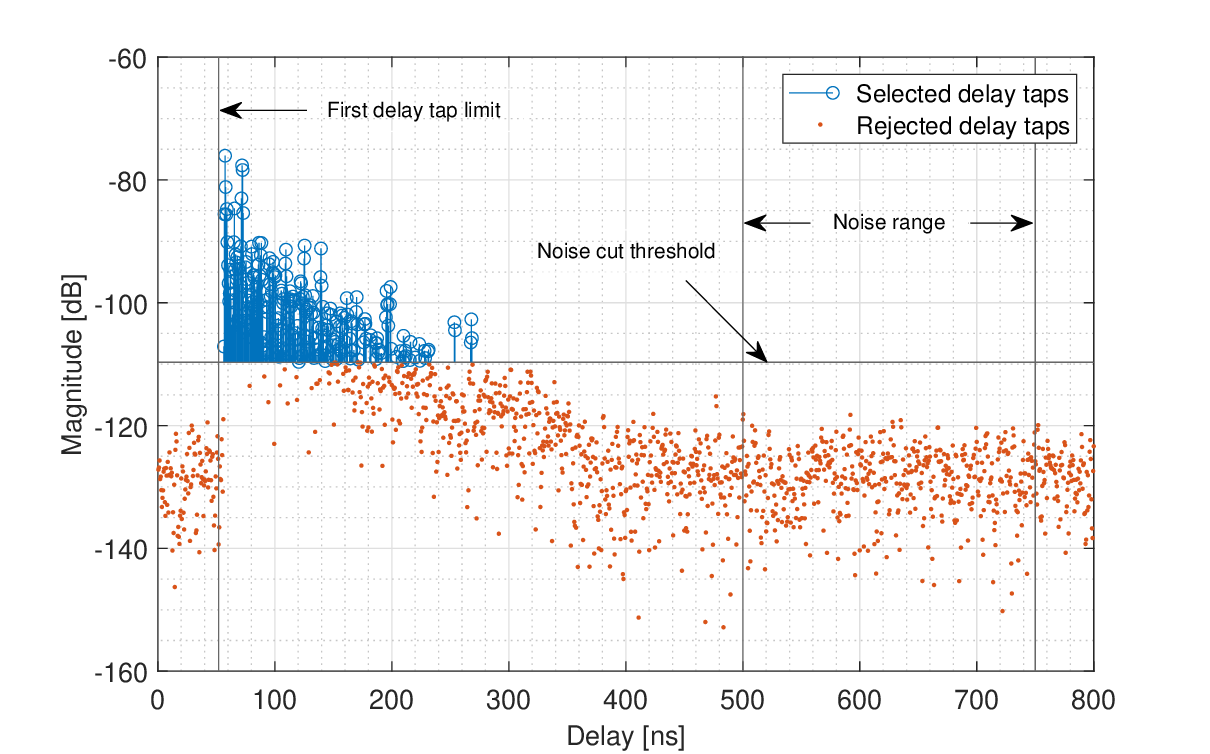}
			\centering
			\caption{Example PDP describing the noise elimination procedure}
			\label{fig:cir_idx_example}
		\end{figure}

		Captured frequency responses of the propagation channel were transferred into time domain channel impulse responses (CIRs) in post-processing using the Hanning window function and the inverse Fourier transformation. The noise level was estimated from time domain channel impulse responses by first identifying the noise range in late delay bins where no contribution from true propagation paths were expected. The noise-cutting threshold was determined by fitting a normal distribution to real and imaginary components of samples within the identified noise range and taking the limit as four times the estimated standard deviation range upward. The taps with a longer delay than those in the noise range were also omitted because it would not be realistic to expect so high delays in this measurement scenario.  This would also eliminate the rising tail effect in the power delay profile resulting from the nature of the Fourier transform. The first delay tap was selected by identifying the 10 dB range from the max power delay tap and selecting the earliest tap within 5~ns in this range of taps. These aspects are visualized in Fig. \ref{fig:cir_idx_example}.
		
		\subsection{Estimating Propagation Channel Parameters}
		
\begin{figure*}
			\centering
			\subfloat[]{\includegraphics[width=0.5\columnwidth]{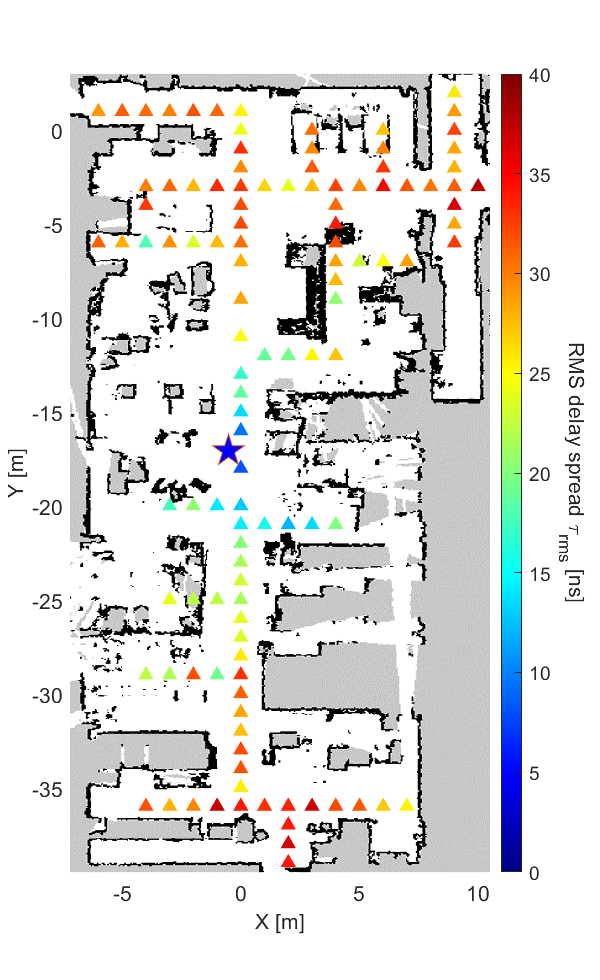}%
				\label{fig:BS0_10_12_DS}}
			\subfloat[]{\includegraphics[width=0.5\columnwidth]{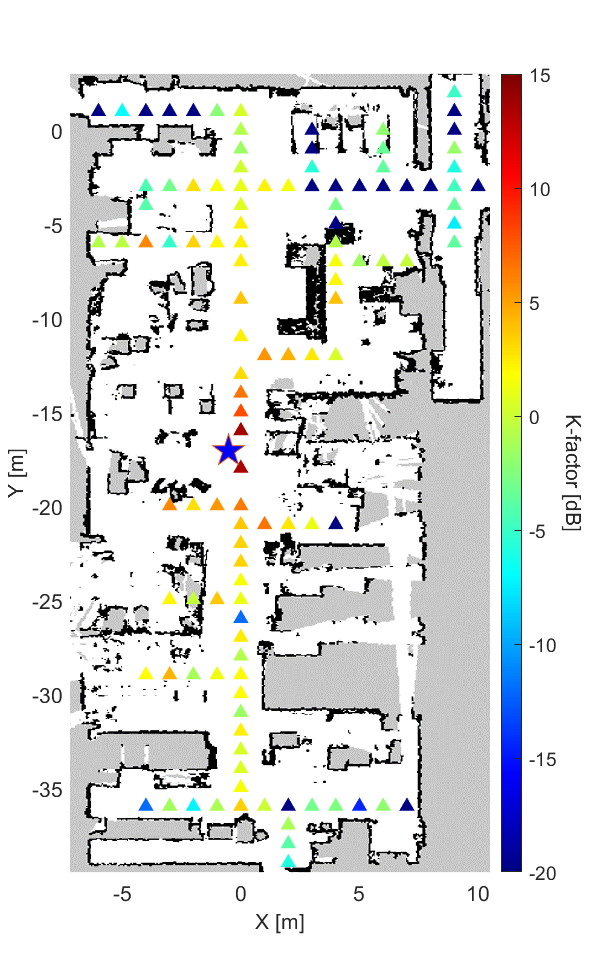}%
				\label{fig:BS0_10_12_K}}
			\subfloat[]{\includegraphics[width=0.5\columnwidth]{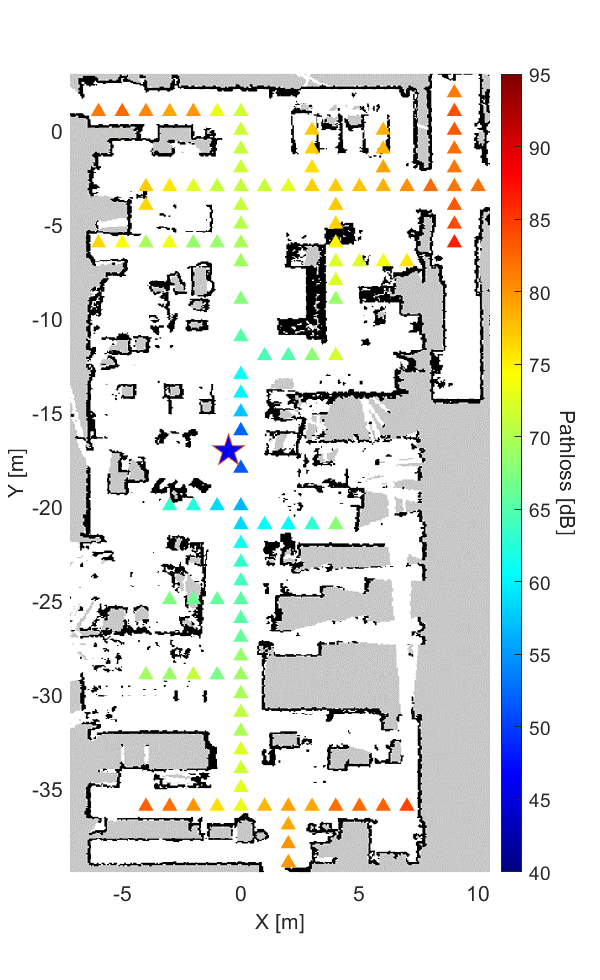}%
				\label{fig:BS0_10_12_PL}}
			\subfloat[]{\includegraphics[width=0.5\columnwidth]{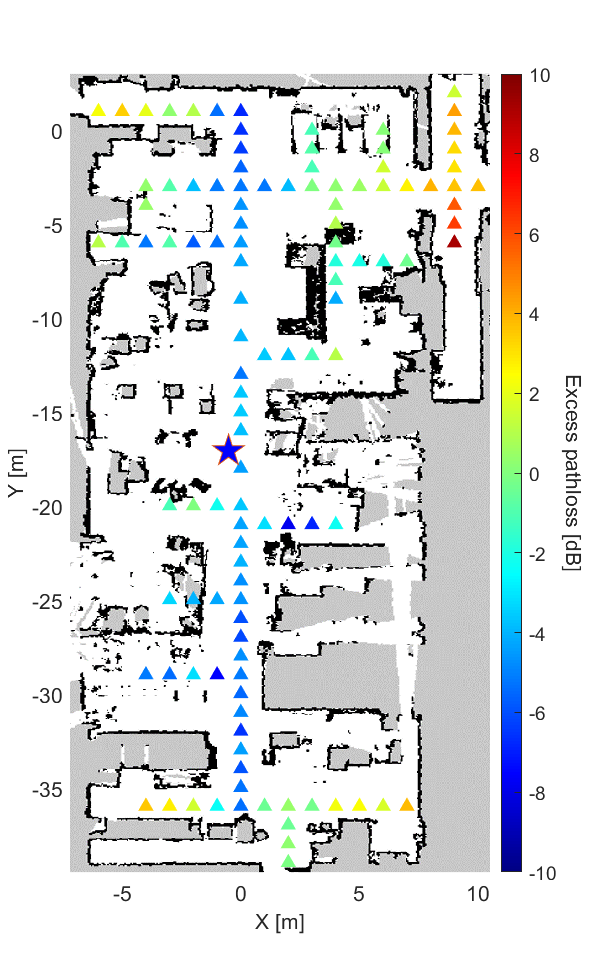}%
				\label{fig:BS0_10_12_EPL}}
			\caption{Large-scale channel characteristics extracted from measurements: (a) delay spread, (b) K-factor, (c) pathloss, and (d) excess pathloss}
			\label{fig:measurement_map_all}
		\end{figure*}

		The so-called large-scale channel parameters: PDP, delay spread, Rician K-factor, pathloss, and excess pathloss were calculated based on denoised omni-directional impulse responses. PDP is defined by calculating the power of each identified delay tap of multipaths. RMS delay spread is calculated as
		
		\begin{equation}
			\sigma_{\tau_{rms}} = \sqrt{\frac{\int_0^\infty{(\tau-\mu_{T_m})^2}A_c(\tau)d\tau}{\int_0^\infty{A_c(\tau)d\tau}}
				\label{rms_delay_spread}}
		\end{equation}
		
		\noindent by first calculating the mean delay spread as
		
		\begin{equation}
			\mu_{T_m} = \frac{\int_0^\infty{\tau}A_c(\tau)d\tau}{\int_0^\infty{A_c(\tau)d\tau}},
			\label{mean_delay_spread}
		\end{equation}
		
		\noindent where \begin{math} \mu_{T_m}(\tau) \end{math} and \begin{math} A_c \end{math} is the mean delay and channel auto-correlation function (power-delay profile). The resulting RMS delay spreads for one of the three TX locations are shown in Fig. \ref{fig:BS0_10_12_DS}, where the color of the RX marker indicates the delay spread. From the figure, we can observe the general trend found many times in prior studies; the delay spread increases with increasing link distance. The range of delay spreads spans from 6.5 to 41.5~ns. The mean values for the factory scenario (InF) in the 3GPP model \cite{3GPP38901} are 26.7~ns and 30.8~ns in the LOS and NLOS cases, respectively (assuming 5~m height for the hall). These are quite well aligned with our observations and the mean values in Tables \ref{table_LOS} and \ref{table_NLOS}.
		
		Rician K-factors were calculated from raw frequency transfer functions using the method of moments introduced by Greenstein \cite{Greenstein1999}. The variation of squared magnitudes was considered over the measured frequency samples since no temporal or spatial sampling was available. Estimated K-factors range between --20 and 8.3~dB, and they are illustrated in Fig. \ref{fig:BS0_10_12_K} for one of the TX locations and corresponding RX locations. It is noteworthy that we observe negative K-factor values in NLOS channel conditions. The negative K-factor values indicate that the scattered components dominate over the direct path component resembling Rayleigh fading.
		
		The measured pathloss was estimated by calculating the mean received power over the measured frequency transfer function and subtracting the antenna gain. Pathloss results are shown in Fig. \ref{fig:BS0_10_12_PL} and they are within 76...44~dB range of values. 
		The measured pathloss was also compared to the theoretical free-space pathloss to determine the excess pathloss, in decibel units $L_\mathrm{excess}=L_\mathrm{meas}-L_\mathrm{free}$. Determined excess path losses are depicted in Fig. \ref{fig:BS0_10_12_EPL} in values between --4.8 and 8.6~dB.
		
		
		The mentioned large-scale parameters were calculated for each link, in total for 254 TX and RX pairs. Then both the mean and standard deviation were calculated per parameter, separately for the sets of LOS and NLOS links. In the case of delay spread the mean and standard deviation are calculated in nanosecond units and for the rest of the parameters in decibel units. Results for LOS and NLOS links are given in Table \ref{table_LOS} and \ref{table_NLOS}, respectively.
		
		
		\begin{table}
			\caption{Statistical parameters LOS links}
			\label{table_LOS}
			\setlength{\tabcolsep}{9.5pt}
			\begin{tabular}{|c|c|c|c|c|}
				\hline
				&
				Delay spread &
				K-factor &
				Excess Pathloss \\
				&
				[ns] &
				[dB] &
				[dB]\\
				\hline
				$\mu$ &
				22.5 &
				2.0 &
				-1.4 \\
				$\sigma$ &
				7.4 &
				1.9 &
				1.0 \\
				\hline
			\end{tabular}
		\end{table}
		
		\begin{table}
			
			\caption{Statistical parameters NLOS links}
			\label{table_NLOS}
			\setlength{\tabcolsep}{9.5pt}
			\begin{tabular}{|c|c|c|c|c|}
				\hline
				&
				Delay spread &
				K-factor &
				Excess Pathloss\\
				&
				[ns] &
				[dB] &
				[dB]\\
				\hline
				$\mu$ &
				29.7 &
				-8.0 &
				0.8 \\
				$\sigma$ &
				6.9 &
				8.5 &
				2.6 \\
				\hline
			\end{tabular}
		\end{table}
		
		\subsection{Comparing measured large scale parameters against 3GPP InF standard channel model}
		Large scale parameters, RMS delay spread, Rician K-factor, and pathloss, were benchmarked against the 3GPP InF standard channel model introduced in \cite{3GPP38901}. The measured large-scale parameters from all TX-RX locations were added together and empirical CDF curves were calculated for all parameters to be compared with corresponding CDFs generated from 3GPP InF standard channel model.
		
		\subsubsection{Delay spread}
\begin{figure}
			\includegraphics[width=0.8\linewidth]{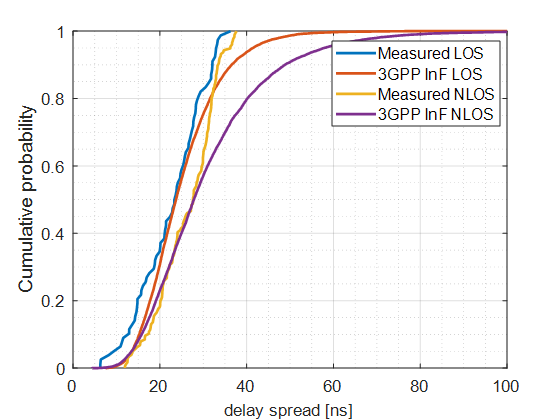}
			\centering
			\caption{LOS and NLOS delay spread comparison between measurements and 3GPP InF model.}
			\label{fig:FR3_LOS_NLOS_delaySpreadCDF}
		\end{figure}

		CDF curves for the 3GPP InF benchmark model were calculated by estimating the factory hall volume and floor, walls and ceiling areas and using the 3GPP InF model formula for delay spread introduced in \cite{3GPP38901} (Table 7.5-6 Part 3). The industrial hall was estimated to be (41m L $\times$ 17m W $\times$ 4m H). Using the mean and standard deviation a CDF curve was calculated with 10'000 randomly generated values for the 3GPP InF delay spread. It is noteworthy that the 3GPP model equation does not depend on the center frequency of the signal.  CDF curve of the 3GPP InF benchmark model was then plotted in Fig. \ref{fig:FR3_LOS_NLOS_delaySpreadCDF} along with the empirical CDF calculated from measured delay spread values for each RX-TX position.
		
		Delay spreads are close to each other in median range values, but for the lower and higher end densities, the 3GPP InF model has higher delay spread values for the LOS case. For the NLOS case, the lower end and median range values are very close to each other in values while in the higher range, the 3GPP model values are significantly higher (Fig. \ref{fig:FR3_LOS_NLOS_delaySpreadCDF}). The noticeable difference in delay spread values in the lowest and uppermost extreme densities is mostly due to the much larger amount of normally distributed random data generated for the 3GPP InF standard model when compared to the amount of measurement data available for this campaign. 
		
		\subsubsection{Rician K-factor}
		
\begin{figure}[]
			\includegraphics[width=0.8\linewidth]{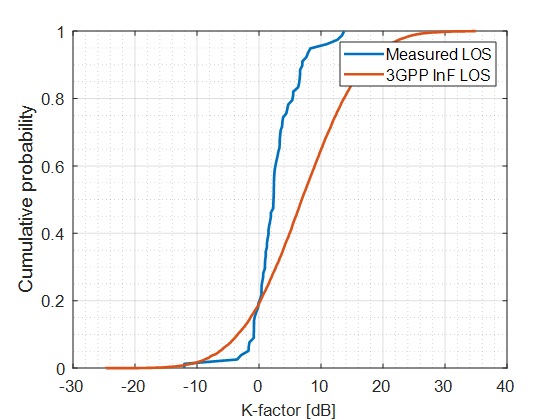}
			\centering
			\caption{K-factor comparison between measurements and 3GPP InF model.}
			\label{fig:FR3_LOS_K-factorCDF}
		\end{figure}

		CDF curves for the 3GPP InF standard channel model were calculated according to \cite{3GPP38901} (Table 7.5-6 Part 3). We generated normally distributed 10k random values with the mean of 7 dB and a standard deviation of 8 dB and calculated a CDF curve from this data. This CDF curve was then plotted in Fig. \ref{fig:FR3_LOS_K-factorCDF} against the empirical CDF curve calculated from measured K-factor values.
		
		When inspecting the CDF curves for the K-factor it can be seen, that the variance in measured K-factor values is significantly smaller than in the case of randomly generated 3GPP InF model data. Even in the median density, the generated 3GPP InF data shows some 5 dB or higher K-factor values than the measured data. Overall the measured K-factor values are more tightly packed closer to the 0 dB range than the reference model, clearly stating that the Rician characteristic of the overall fading is less evident in this measurement campaign than the 3GPP standard channel model would expect. A typical condition for Rician fading channel is a single dominant path and several weaker ones. A possible explanation for the lower-than-expected K-factor in measured data is the presence of several strong reflected paths from metallic objects. This is observable in the example PDP of a LOS case in Fig. \ref{fig:cir_idx_example} with only 2.5~dB attenuated reflected paths.
		
		\subsubsection{Pathloss}
		The measured and 3GPP pathlosses were compared by calculating the pathloss for each RX-TX pair link distance and then comparing the results against 3GPP InF-DL and InF-SL pathloss models. Inf-DL is a reference pathloss model for densely cluttered indoor factories and both the RX and TX heights are low compared to the clutter. InF-SL on the other hand is a reference model for sparsely cluttered indoor factories with low RX and TX heights.
		
		The measured pathloss was calculated by taking the mean power of the Hanning window filtered frequency responses and subtracting the antenna gains from the result. The 3GPP InF-DL and InF-SL pathloss for each link was calculated using formulas specified in \cite{3GPP38901} (Table 7.4.1-1).
		
\begin{figure}[]
			\includegraphics[width=0.8\linewidth]{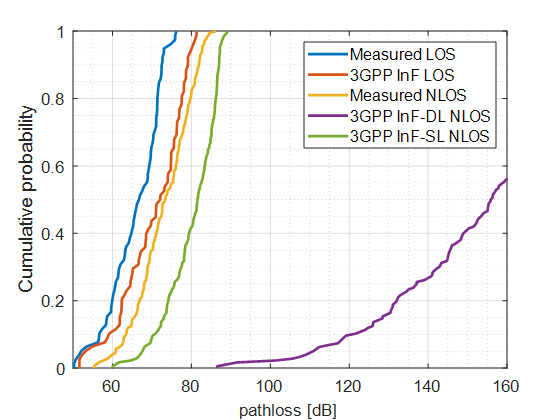}
			\centering
			\caption{LOS and NLOS pathloss comparison between measurements and 3GPP Inf-DL and Inf-SL models.}
			\label{fig:FR3_LOS_NLOS_Inf-DL_SL_pathLossCDF}
		\end{figure}

		Empirical CDF curves were calculated from both measured pathloss values and 3GPP InF-DL and InF-SL reference models and plotted for comparison in Fig. \ref{fig:FR3_LOS_NLOS_Inf-DL_SL_pathLossCDF}. For LOS condition InF-SL and Inf-DL models are equal, and it can be seen that the measured and reference model values are close to each other in the lower range but they depart as we move to the higher pathloss values. The difference is between 10 - 15 dB at maximum but overall we can observe that the measured pathloss values are lower than the 3GPP InF standard model. 
		
		For the NLOS paths, we can see that the densely cluttered pathloss reference model from 3GPP does not match the measured pathloss values, which is most likely due to the large open spaces in the production hall (Fig. \ref{fig:EnvPhoto}). The differences in this case are in the order of magnitude of 100 dB. The sparsely cluttered reference model matches the measured data better. 
		
				\section{Conclusion}
				We have analyzed propagation transfer functions measured in a factory-like production lab by Aalborg University at 10--12 GHz radio frequency using a vector network analyzer and omni-directional antennas. The objective was to study radio propagation in industrial environments at the so-called FR3 in short-range communication scenarios.
				
				We found general trends of increasing delay spread, Rician K-factor, and path loss with increasing link distance. This is not a surprise, since the last one is evident by definition and the two former observations are well aligned with similar studies on various radio frequencies. Delay spreads range between 6.5 and 41.5~ns and K-factors between --20 and 8.3~dB. The corresponding mean values are 22.5~ns and 2~dB, and 29.7~ns and -8~dB in LOS and NLOS conditions, respectively.
				
				We compared measured large-scale parameters: delay spread, Rician K-factor, and pathloss against their corresponding 3GPP Indoor factory reference models by comparing empirical CDFs. The measured delay spreads were found to match the reference model in the median range densities. The measured K-factors were overall found to be quite low, reaching 10 dB only in the most extreme cases. This indicates that even in the clear LOS condition, the LOS component does not dominate. Instead, there are several strong paths in addition to the LOS path. It can be presumed that the strongly reflecting metallic structures in the industrial hall provide a rich delay spread without significantly attenuating the gain of these delayed paths. Therefore, the power ratio of the strongest LOS paths remains close to those of the reflected paths. The reference model K-factors were found to be overall some 5 -- 10 dB higher than the measured values. 
				
				The measured pathlosses were found to be significantly lower than those of the 3GPP reference model for the \emph{densely cluttered factory}, indicating that the densely cluttered scenario does not characterize well the measured environment, which had large open spaces and corridors. With the \emph{sparsely cluttered} reference model, the difference compared to the measured values was smaller but still some 10 dB. Therefore, the current 3GPP InF path loss models may be too pessimistic for the mid-band propagation in industrial environments.
				
				As a standardization activity, the 3GPP RAN1 WG1 has an ongoing study item for defining channel models for FR3 (7--24)\,GHz band, and the findings from this indoor factory measurement campaign have been recently reported in it \cite{RAN1}.
				
				\section*{Acknowledgement}
				This work was supported in part by the EU SNS JU 6G-SHINE project (Grant Agreement No. 101095738). Aalborg University would like to thank Anritsu for providing measurement equipment and for their funding of the  measurement campaigns. The authors would like to thank Prof. Enrico Maria Vitucci from CNIT, Italy for reviewing the manuscript.

			\end{document}